\documentclass[aps,prb,reprint,nopacs,superscriptaddress,longbibliography]{revtex4-1}
 \usepackage[colorlinks=true, citecolor=blue, urlcolor=blue, linkcolor=blue]{hyperref} 
\usepackage{epsfig,mathrsfs,color,latexsym,subfigure,marginnote,graphicx,verbatim,relsize,mathrsfs,color,array,amsmath,amsfonts,amssymb,graphicx}
\renewcommand{\BibitemShut}[1]{}

\begin{document}

\title{K$_2$CoS$_2$: A new two-dimensional in-plane antiferromagnetic insulator}

\author{Anan Bari Sarkar}
\affiliation{Department of Physics, Indian Institute of Technology, Kanpur 208016, India}
	
\author{Barun Ghosh}
\affiliation{Department of Physics, Indian Institute of Technology, Kanpur 208016, India}
	
\author{Bahadur Singh} 
\email {bahadursingh24@gmail.com}
\affiliation{Department of Physics, Northeastern University, Boston, Massachusetts 02115, USA}

\author{Somnath Bhowmick}
\affiliation{Department of Materials Science and Engineering, Indian Institute of Technology, Kanpur 208016, India}
	
\author{Hsin Lin}
\affiliation{Institute of Physics, Academia Sinica, Taipei 11529, Taiwan}

\author{Arun Bansil}
\affiliation{Department of Physics, Northeastern University, Boston, Massachusetts 02115, USA}		
	
\author {Amit Agarwal}
\email{amitag@iitk.ac.in}
\affiliation{Department of Physics, Indian Institute of Technology, Kanpur 208016, India}
	
\date{\today}

\begin{abstract} 
Recent discovery of two-dimensional (2D) magnetic materials has brought magnetism to the flatland and opened up exciting opportunities for the exploration of fundamental physics as well as novel device applications. Here, we predict a new thermodynamically stable 2D magnetic material, K$_2$CoS$_2$, which retains its in-plane bulk antiferromagnetic (AFM) order down to the monolayer and bilayer limits. Magnetic moments ($2.5 \mu_B/$Co) are found to form a quasi-one-dimensional anti-ferromagnetically ordered chain of Co-atoms. The non-magnetic electronic spectrum of the monolayer film is found to host flat bands and van-Hove singularities, which play a key role in stabilizing the magnetic ground state. Based on classical Monte-Carlo simulations, we estimate the Neel temperature for the AFM monolayer to be $\approx 15$K. Our study demonstrates that K$_2$CoS$_2$ hosts a robust AFM state which persists from the monolayer limit to the bulk material.
\end{abstract}

\maketitle
\nopagebreak

\section{Introduction}
Discovery of magnetism in two-dimensional (2D) materials has opened up exciting possibilities not only for exploring fundamental physics of magnetism in lower dimensions but also their potential use as novel platforms for energy storage and quantum information processing applications\cite{huang2017,gong2017,burch2018,cortie2019,samarth2017,huang2018,jiang2018,burch2018a,xing2017,seyler2018}. Magnetic ordering in spin-rotation symmetric 2D systems is generally suppressed by thermal and quantum fluctuations, preventing magnetism in 2D\cite{mermin1966}. Magnetic order in 2D can be stabilized, however, via anisotropic interactions. One of the earliest theoretical examples of 2D magnetism dates back to the exact solution of the 2D Ising model in 1944\cite{onsager1944}. On the experimental side, the magnetism of thin films has a long history\cite{bell2000,cortie2019}, but the field has seen recent revival following the realization of magnetism in monolayer CrI$_3$\cite{huang2017} and Cr$_2$Ge$_2$Te$_6$\cite{gong2017} films. While monolayer CrI$_3$ is found to be an Ising-type ferromagnet (below $45$ K), bilayer Cr$_2$Ge$_2$Te$_6$ is described by a Heisenberg spin model with additional anisotropic terms. In both these materials, the interlayer magnetic ordering is antiferromagnetic (AFM). Many other layered materials have been predicted to support magnetism in their thin-film limit. Notable examples include FePS$_3$ (Ising-type anti-ferromagnet)\cite{wang2016,lee2016}, MnSe$_2$\cite{ohara2018}, Fe$_3$GeTe$_2$ \cite{Liu2017}, transition metal tri-halides (AX$_3$)\cite{liu2016,zhang2015,he2016,zhou2018,he2017,sun2018,iyikanat2018,sarikurt2018,weber2016,sheng2017,tomar2019}, transition metal di-halides (AX$_2$) \cite{kan2014,zhuang2016,wang2018,botana2019} and MXenes \cite{khazaei2012,zhang2017,he2016a,kumar2017}.

In this paper, we predict K$_2$CoS$_2$ to be a new thermodynamically stable magnetic material with an in-plane AFM ground state down to the monolayer limit. Our density-functional-theory (DFT) based calculations reveal that K$_2$CoS$_2$ hosts flat bands and Van-Hove singularities (VHSs) near the Fermi energy in its nonmagnetic band structure. These features of the electronic structure coupled with the strong Coulomb interaction arising from the Co-$d$ orbitals stabilizes the AFM state. Remarkably, bulk K$_2$CoS$_2$ hosts an in-plane AFM ordering with strongly coupled quasi-one-dimensional (1D) AFM Co chains. This magnetic ordering is preserved down to thin films and the monolayer limit. We show that the magnetism of monolayer K$_2$CoS$_2$ is well described by a 2D Heisenberg model with an onsite anisotropy term with an exchange coupling that is much stronger along the quasi-1D Co-chains compared to the inter-chain coupling. Based on our Monte-Carlo calculations, we predict the Neel transition temperature for the monolayer to be $T_N \approx 15$ K, which is comparable to the experimentally reported value of $10$ K in bulk K$_2$CoS$_2$\cite{bronger1990}. Our study demonstrates for the first time that monolayer K$_2$CoS$_2$ could provide a robust material platform for exploring 2D magnets with strongly coupled quasi-1D AFM chains. 

The paper is organized as follows. Sec. \ref{Comp} presents computational details. In Sec.~\ref{bulk}, we discuss the electronic properties and magnetic ordering of bulk K$_2$CoS$_2$. The magnetic and electronic properties of the monolayer K$_2$CoS$_2$ are explored in Sec.~\ref{monolayer}. In Sec.~\ref{bilayer}, we briefly discuss the magnetic and electronic structures of bilayer K$_2$CoS$_2$. Finally, we summarize the findings of our study in Sec.~\ref{conclusion}.

\section{Computational Details} \label{Comp}

We performed {\it ab-initio} calculations within the framework of the DFT using the Vienna {\it ab-initio} simulation package (VASP)\cite{kresse1996efficient, kresse1999ultrasoft}. Exchange-correlation effects were treated within the generalized-gradient-approximation (GGA)\cite{Perdew1996}. Since the GGA often fails to correctly describe localized electrons\cite{torelli2019high}, we considered an onsite Coulomb interaction for Co-$d$ orbitals within the GGA+$U$ scheme with $U_{eff}= 4.4$ eV. This value of $U_{eff}$ was obtained self-consistently using the linear-response theory as discussed in Appendix \ref{hubbard}. The robustness of electronic and magnetic properties was confirmed by varying the value of $U_{eff}$ from 0 to 6 eV. The kinetic energy cut-off used for the plane-wave basis set was 600 eV. A $\Gamma$-centered $22\times22\times 1$ ($22\times22\times 8$) $k$-mesh\cite{Monkhorst1976} was used to perform Brillouin zone (BZ) integrations for the monolayer/bilayer (bulk) K$_2$CoS$_2$. Thin-film calculations were carried out using a slab model with a vacuum layer of 16 \AA~to avoid interactions between the periodically repeated images. The in-plane lattice constants and atomic positions were optimized until the residual force on each atom became less than 10$^{-3}$ eV/\AA. Dynamical stability of the monolayer was confirmed via phonon calculations using the Phonopy code with a 3 $\times$ 3 $\times$ 1 supercell\cite{phonopy}. 
Thermal stability of the monolayer was analyzed by performing  {\it ab-initio} molecular dynamics simulations in a canonical ensemble with a time step of 0.5 fs for 10,000 steps at 400 K.

\section{Bulk $\mathbf {K_2CoS_2}$}\label{bulk}

\begin{figure}[t!]
\includegraphics[width=0.99\linewidth]{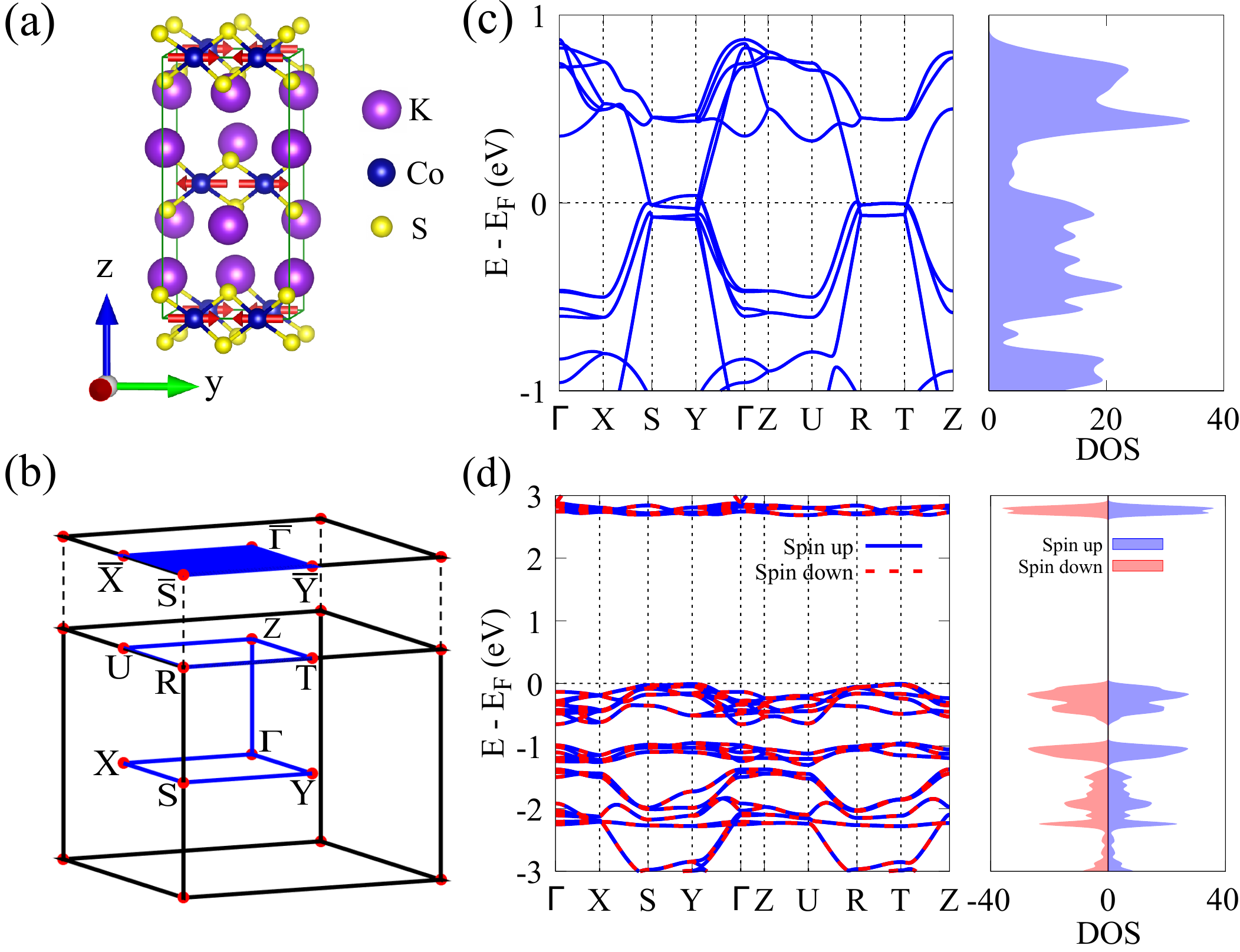} 
\caption{(a) The layered crystal structure of bulk K$_2$CoS$_2$ in the magnetic ground state. Magnetic moments of the Co atoms form an AFM chain-like configuration. (b) Bulk Brillouin zone (BZ) and the associated surface BZ projected on to the (001) surface. High-symmetry points are marked. The calculated band structure and total density of states of bulk K$_2$CoS$_2$ with $U_{eff} = 4.4$ eV for (c) nonmagnetic and (d) AFM state.} 
\label{fig1}
\end{figure}

Bulk K$_2$CoS$_2$ is a ternary chalcogenide that crystallizes in the orthorhombic space group $Ibam$ (No. 72). The crystal structure is layered, with the individual K$_2$CoS$_2$ layers extended in the $xy$ plane (see Fig.~\ref{fig1}(a)) and held together by weak inter-plane van der Waals forces. Magnetism originates from Co atoms which form an edge-sharing tetrahedral arrangement as seen in Fig.~\ref{fig1}(a). In the AFM state, we find the Co-S bonds (bond length $\sim$ 2.37\AA) to be stronger than the K-S bonds (bond-length $\sim$ 3.31\AA). The relaxed lattice parameters in the nonmagnetic (NM), ferromagnetic (FM), and AFM configurations are listed in Table~\ref{tab11}. Our theoretically predicted ground-state bond lengths and lattice constants for the bulk AFM phase are consistent with the corresponding experimental values \cite{bronger1990}. Interestingly, we find the volume of the bulk unit cell for magnetic configurations to be larger than for the nonmagnetic case by $\sim 10$\% (Table~\ref{tab11}). This magneto-volume effect\cite{takahashi2013spin} can be used to obtain an estimate of the magnetic ordering temperature \cite{ran2016phase}.

\begin{table}[t!]
\caption{Optimized lattice parameters and the energy per Co atom (E) for the nonmagnetic (NM), antiferromagnetic (AFM), and ferromagnetic (FM) configurations of bulk K$_2$CoS$_2$. The ground state is the in-plane AFM configuration. Experimental lattice parameters\cite{bronger1990} in the AFM state are listed.} 
\centering  
\resizebox{0.99\columnwidth}{!}{
\begin{tabular}{c  c  c  c  c  c}
		\hline \hline
		Configuration & \hspace{.1cm} $a$ (\AA) \hspace{.1cm} & \hspace{.1cm} $b$ (\AA) \hspace{.1cm} & \hspace{.1cm} $c$ (\AA) \hspace{.1cm} & Volume (\AA$^3$) \hspace{.1cm} &  E (eV) \\ 
		\hline \hline
		 NM & 6.625 &  6.006 & 12.349  & 491.495 & -17.246\\
		 \hline
		 AFM & 6.821  & 6.184  & 12.714  & 536.425 & -19.423 \\
		 \hline
		 FM & 6.835 & 6.196 & 12.739 & 539.564 & -19.357 \\
		 \hline
		 AFM$_{exp}$ & 6.710 & 6.085 & 12.491 & 510.012 & --\\
		\hline 
\end{tabular} }
\label{tab11}
\end{table}

Our GGA+$U$-based ground state of bulk K$_2$CoS$_2$ displays an in-plane AFM ordering of the Co magnetic moments, see Appendices~\ref{hubbard} and \ref{energy} for details. In order to identify the magnetic easy axis, we have computed total energies including spin-orbit coupling (SOC) for the AFM and FM configurations in which Co spins are aligned along the $x$, $y$, or the $z$ axis. The energy is found to be minimum when the moments are aligned antiferromagnetically along the Co chains ($y$ direction in Fig. \ref{fig1}(a)). These results are consistent with the magnetic configuration observed in neutron diffraction experiments\cite{bronger1990}. The computed magnetic moment of Co atoms is $\sim$2.5 $\mu_B$, which is in agreement with the experimentally reported value of 2.5 $\mu_B$ \cite{bronger1990}. 

The nonmagnetic band structure of bulk K$_2$CoS$_2$ is shown in Fig.~\ref{fig1}(c). It is seen to be metallic with several band-crossings at the Fermi level. A striking feature is the presence of multiple `flat' bands near the Fermi level along the $S-Y$ and $R-T$ directions and the associated VHSs in the density of states (DOS) [right panel of Fig.~\ref{fig1}(c)], which could drive many-body instabilities in the lattice, charge, and spin channels\cite{PhysRevB.56.3159,PhysRevLett.78.1343,PhysRevB.35.3359}. These VHSs could also stabilize the AFM state in K$_2$CoS$_2$.

Figure ~\ref{fig1}(d) shows the band structure for the AFM magnetic configuration of K$_2$CoS$_2$. It is seen to be insulating with a relatively large bandgap of $\sim 2.69$ eV. Notably, the SOC has a negligible effect on the band structure here although its inclusion is key for breaking the $SU(2)$ spin-rotation symmetry and stabilizing the anisotropic magnetic ground state\cite{guo1991,tung2007}.

\begin{table}[t]
	\caption{Relaxed lattice parameters and the corresponding energies per Co atom (E) of monolayer K$_2$CoS$_2$ for three different magnetic configurations. }
	\centering  
	\centering  
\resizebox{0.99\columnwidth}{!}{
	\begin{tabular}{ c c c c c }
		\hline \hline
		Configuration & \hspace{0.15cm} $a$ (\AA) \hspace{0.15cm} & \hspace{0.15cm}$b$ (\AA) \hspace{0.15cm}& \hspace{0.15cm}Area (\AA$^2$) \hspace{0.15cm} & \hspace{0.15cm} E (eV) \\ 
		\hline \hline 
		NM & 6.448 &  5.846  &  37.695 & -16.798\\
		\hline
		AFM & 6.675 & 6.051  & 40.390 & -18.914\\
		\hline
		FM & 6.684 & 6.059   & 40.498 & -18.849 \\
		\hline 
	\end{tabular}}\\
	\label{tab2} 
\end{table}

\section{Monolayer $\mathbf {K_2CoS_2}$}\label{monolayer}

We now turn to discuss the electronic and magnetic properties of monolayer K$_2$CoS$_2$ using the structure obtained by separating one layer of K$_2$CoS$_2$ from the bulk. 
Notably, our calculated exfoliation energy (EE) is 0.64 J/m$^2$, which is comparable to the EEs of well-known 2D materials such as graphene (0.32 J/m$^2$)\cite{PhysRevB.76.155425}, MoS$_2$ (0.29 J/m$^2$)\cite{PhysRevLett.108.235502}, SnP$_3$ (0.71 J/m$^2$) \cite{Ghosh18a}, GeP$_3$ (1.14 J/m$^2$) and NaSnP (0.81 J/m$^2$).
\footnote{{K$_2$CoS$_2$ belongs to a large family of quasi-1D chain materials, which have been exfoliated as 2D sheets \cite{noguchi2019weak,springborg2003quasi}. Since the exfoliation energy of K$_2$CoS$_2$ monolayer is comparable to that of other well-known 2D materials, it is highly likely that it can also be exfoliated as 2D sheets.}}

\begin{figure}[t!]
\includegraphics[width=0.99\linewidth]{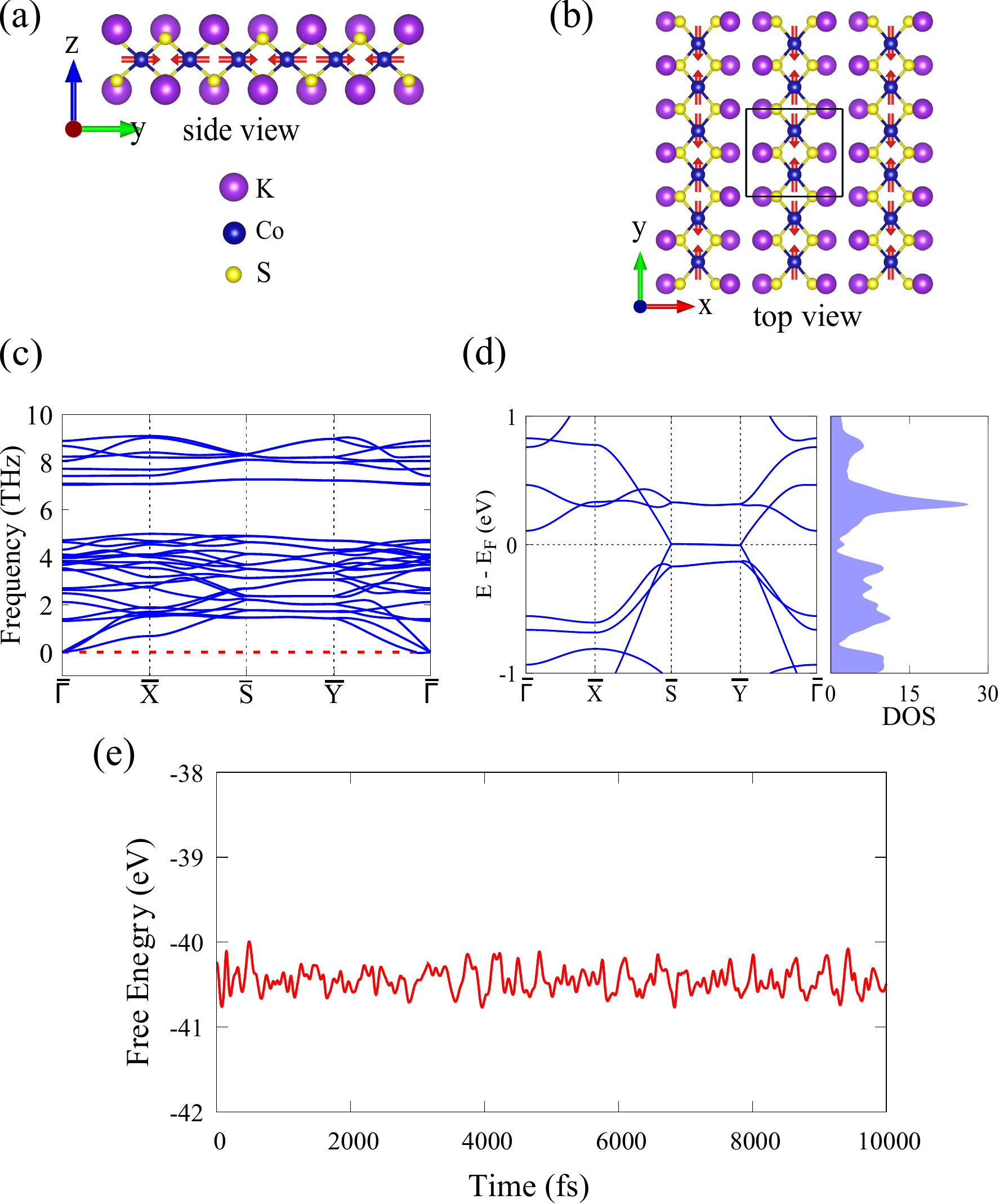} 
\caption{The (a) side and (b) top view of the AFM crystal structure of monolayer K$_2$CoS$_2$. Magnetic moments on Co are aligned along the Co-Co chains in the $y$ direction. (c) Phonon spectrum for monolayer K$_2$CoS$_2$. (d) Nonmagnetic band structure and density of states (DOS). The presence of flat bands and VHSs at the Fermi energy indicates propensity of the system toward forming correlated magnetic states. (e) Total free energy of monolayer K$_2$CoS$_2$ during {\it ab-initio} molecular dynamics simulation at 400 K.
}
\label{fig2}
\end{figure}

The crystal structure of monolayer in the AFM configuration is shown in Figs.~\ref{fig2}(a) and (b). The Co atoms are arranged in quasi-1D chains that extend along the $y$ direction. The Co-S and K-S bond lengths are 2.37~\AA~and 3.20~\AA, respectively. The relaxed lattice parameters for the three different magnetic configurations considered are listed in Table~\ref{tab2}. The unit-cell area in the magnetic configurations is larger than the nonmagnetic case, indicating the presence of a magneto-volume-like effect similar to the bulk material. We confirmed the dynamical stability of the monolayer by computing the phonon spectrum, shown in Fig. \ref{fig2}(c), where no imaginary frequency was found.
We have also carried out an {\it ab-initio} molecular dynamics simulation at 400 K to check the thermal stability of the monolayer. Variation of the free energy as a function of simulation time is shown in Fig. \ref{fig2}(e). The monolayer structure was found to remain intact at the end of the simulation, indicating thermal stability of the monolayer.

Band structure and DOS for the nonmagnetic configuration of monolayer are shown in Fig.~\ref{fig2}(d). Two `Dirac like' linearly-dispersing bands along the $\bar{X}-\bar{S}$ and $\bar{\Gamma}-\bar{Y}$ directions are seen to cross at the $\bar{S}$ and $\bar{Y}$ points. These bands remain degenerate and almost flat along the entire $\bar{S}-\bar{Y}$ line. The bandwidth of these degenerate bands is small (10 meV along $\bar{S}-\bar{Y}$ direction), which leads to the VHSs at the Fermi level. Like the bulk nonmagnetic system, the presence of flat bands and the associated VHSs here suggests the possibility of many-body quantum instabilities including magnetism through the Stoner criterion in monolayer K$_2$CoS$_2$.

\begin{figure*}[ht]
\centering
\includegraphics[width=0.9\linewidth]{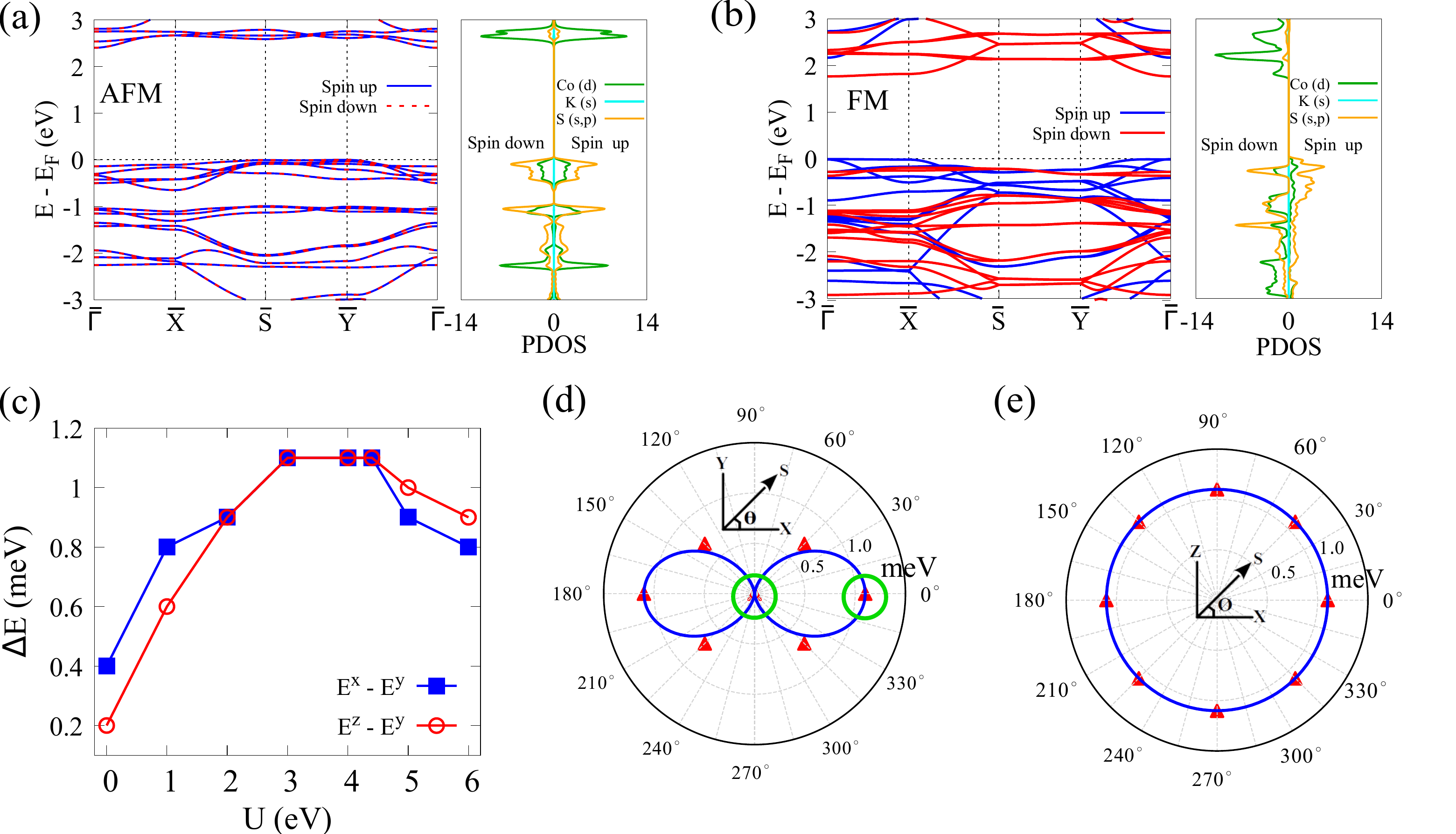} 
\caption{Band structures and partial density of states in monolayer K$_2$CoS$_2$ for (a) AFM and (b) FM configurations. (c) Relative energy difference per unit cell with respect to the energy ($E^y$) of the lowest energy configuration over a range of $U$ values for different spin orientations (denoted by superscripts) in the AFM configuration. (d) and (e) Calculated magnetic anisotropy energy (MAE) per unit cell when spin direction is rotated in the $xy$ (d) and $xz$ (e) planes in the AFM configuration for $U$ = 4.4 eV. Solid blue lines give results obtained from the effective spin Hamiltonian of Eq.~\ref{HSM}, {\it i.e.,} MAE($\theta, \phi$) = E$^{\theta, \phi}_{AFM}$ - E$^{y}_{AFM}$. Red triangular points give DFT-based values of MAE. Data points enclosed by the green circles in (d) were used to calculate the anisotropic exchange parameters in the effective Hamiltonian of Eq.~\ref{HSM}.}
\label{fig3}
\end{figure*}

In order to gain further insight into the nature of the magnetic ground state, we have computed energies of K$_2$CoS$_2$ monolayer where the magnetic moments are constrained to be oriented along different directions, see Appendices \ref{hubbard} and \ref{energy} for details. Interestingly, the magnetic ground state of the monolayer is the same as that of bulk K$_2$CoS$_2$ where Co magnetic moments are ordered anti-ferromagnetically along the Co chains. The robustness of these results was further checked by calculating the ground state energy with an on-site Coulomb potential $U$ added on the Co atoms. The AFM state with moment along $y$ direction was found to be preserved over a large range of $U$ values, see Fig.~\ref{fig3}(c).

Monolayer K$_2$CoS$_2$ is an AFM insulator with a relatively large bandgap of $\sim$2.40 eV  as seen from the band structure of Fig.~\ref{fig3}(a). The band structure contains VHSs and several bands with little energy dispersion. However, unlike the nonmagnetic case, the VHSs here are located quite far from the Fermi energy due to the large size of the insulating bandgap. The states near the Fermi energy are admixtures of Co-$d$ and S-$p$ orbitals as shown by the partial density of states (PDOS) in Fig.~\ref{fig3}(a). 
Note that Co d states are shifted away from the Fermi energy due to the onsite Coulomb repulsion, leaving the states near the Fermi energy to be dominated by S. Magnetic moment per Co atom is 2.5$\mu_B$, which is similar to the bulk value. S atoms are not spin polarized. Band structure of the FM state is shown in Fig.~\ref{fig3}(b). The spin-up and spin-down states are now separated in energy due to FM ordering. However, the insulating state is preserved with a smaller bandgap of $\sim 1.77$ eV compared to the AFM state.

We have considered magnetic anisotropy energy (MAE) in monolayer K$_2$CoS$_2$. For this purpose, we constructed the following minimal quadratic spin Hamiltonian that reasonably captures the magnetic ground state of the monolayer. 
\begin{equation} \label{HSM}
\small H =  -\frac{1}{2} \sum_{\langle ij\rangle}\left[J^xS^x_{i}S^x_{j} + J^yS^y_{i}S^y_{j} + J^zS^z_{i}S^z_{j} \right] - \sum_{i} D (S^y_i)^2~
\end{equation}
Here, $S$ is the spin (magnetic moment = 2.5 $\mu_B$) operator, $i$ denotes the Co sites, and the summation $\langle ij \rangle$ runs over the nearest-neighbor Co ions. $J^a$ with $a =\{x,y,z\}$ denotes the anisotropic spin-exchange interaction energy, and $D$ is the onsite magnetic anisotropy parameter. Values of the various parameters in the Hamiltonian of Eq.~\ref{HSM} were obtained by calculating energies of the K$_2$CoS$_2$ monolayer  in the nonmagnetic as well as different magnetic configurations where the magnetic moments are taken to lie along the $x$, $y$, or $z$ direction.  $J^a$’s and $D$ are then obtained from the energy differences of various magnetic configurations in the FM and AFM states, see appendix \ref{anisotropy} for details. For $U_{eff} = 4.4$ eV, the values are: $\{J^x,J^y,J^z, D \} = \{ -5.16, -5.08, -5.16, 0.17\} $ meV. These parameters also allow us to calculate the MAE per unit cell by rotating the spin in the $xy$ and $xz$ planes as shown in Figs.~\ref{fig3}(d) and (e), respectively. The solid-blue lines in panels (d) and (e) are obtained from Eq.~\ref{HSM}, {\it i.e}, MAE($\theta, \phi$) = $E_{\rm AFM}^{\theta, \phi} - E_{\rm AFM}^{y}$, where $\theta$ and $\phi$ are the angles spin direction makes with the $x$ axis in the $xy$ and $xz$ planes. The red triangular points represent the MAE calculated from first-principles results. Points enclosed by the green circles in Fig.~\ref{fig3}(d) were used to estimate the value of the anisotropy parameter $D$ (0.17 meV). The excellent agreement between the MAE results based on the spin model of Eq.~\ref{HSM} and the DFT calculations show the efficacy of the model Hamiltonian of Eq.~\ref{HSM}.

Notably, our in-plane MAE of $\sim$ 1.10 meV/unit cell is comparable to the MAE values in CrI$_3$ (1.60 meV/unit cell)\cite{yang2019strain} and Fe$_3$P (0.72 meV/Fe)\cite{zheng2019high}, indicating the robustness of finite temperature magnetism in K$_2$CoS$_2$ monolayer. We have calculated the inter-chain exchange integrals between the two nearest Co atoms (in the $x$ direction) on the neighboring Co chains, and found these to be at least two orders of magnitude smaller ($\sim$ 0.01 meV) than the nearest-neighbor $J^a$ values obtained within the Co-chains.   

\begin{figure}[h!]
\includegraphics[width=0.99\columnwidth]{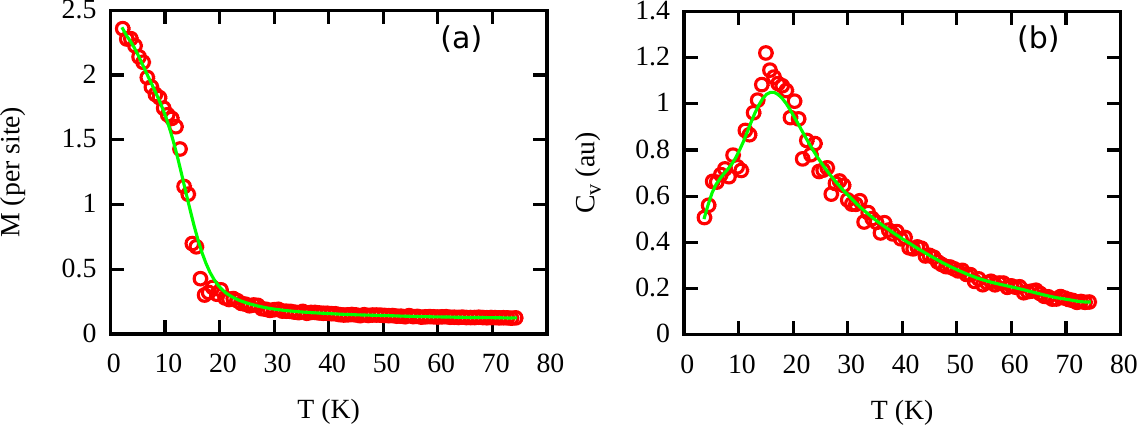} 
\caption{(a) Classical Monte-Carlo simulation results for monolayer K$_2$CoS$_2$ showing the magnitude of the magnetic moment (per Co site) and (b) the corresponding specific heat (per Co site). These results indicate the presence of a magnetic phase transition at $T_N \approx 15 K$ from the AFM to the paramagnetic state.}
\label{fig3a}
\end{figure}

We have estimated the magnetic transition temperature of monolayer K$_2$CoS$_2$ based on the spin model of Eq.~\ref{HSM} in which the next-nearest-neighbor coupling was added. Monte Carlo (MC) simulations were then performed on a $32\times 32$ spin lattice with 2048 spins. Starting with randomly oriented spins, the simulation allowed all spins to rotate freely in three-dimensional space.  At every temperature, $10^7$ steps were used for equilibration, followed by the calculation of the average value over $10^5$ further steps. The magnitude of the total spin (per Co-site) in each of the magnetic sublattices is plotted as a function of temperature in Fig.~\ref{fig3a}(a). We also calculated the temperature dependence of the specific heat, which is given by 
\begin{equation}
C_v\sim \frac{\langle E^2\rangle-\langle E\rangle^2}{T^2}~
\end{equation}
Results for $C_v$ per site are shown in Fig.~\ref{fig3a}(b). Both the sublattice magnetization and the specific heat curves clearly show a magnetic transition from a paramagnetic to an AFM state with decreasing temperature around $T_{N} \approx 15 K$. Our analysis shows that spin interactions in K$_2$CoS$_2$ are dominated by the large intra-Co-chain Heisenberg-like exchange coupling with an onsite spin anisotropy. As a result, the AFM ordering in K$_2$CoS$_2$ is very robust and persists from bulk to monolayer with almost the same Neel temperature. In this connection we also explored the stability and magnetism of a bilayer of K$_2$CoS$_2$ as discussed below.

\section{Bilayer $\mathbf {K_2CoS_2}$} \label{bilayer}

\begin{figure}[t!]
\includegraphics[width=0.99\linewidth]{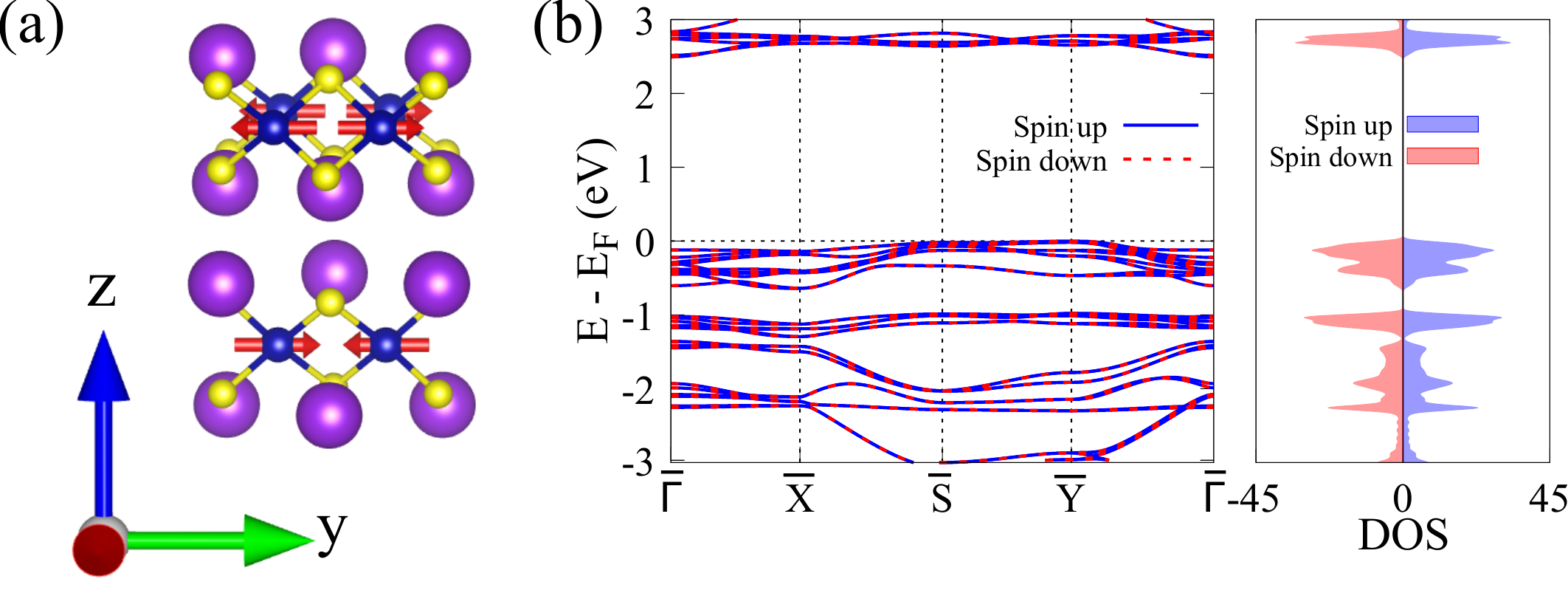} 
\caption{(a) Crystal structure of the K$_2$CoS$_2$ bilayer where the Co atoms are coupled antiferromagnetically. (b) The associated band structure and DOS obtained with $U_{eff}$ = 4.4 eV.}\label{fig4}
\end{figure}

The K$_2$CoS$_2$ bilayer was constructed by stacking two monolayers in the manner in which they are stacked in the bulk structure (Fig.~\ref{fig4}(a)). In particular, the two monolayers in the bilayer structure are shifted with respect to each other by half of the lattice vector along the $x$ direction.  The relaxed lattice parameters for the  magnetic and nonmagnetic configurations are summarized in Table~\ref{tab3}. Like the bulk and the monolayer, the bilayer also assumes a robust AFM ground state and displays the magneto-volume effect.  The band structure and spin-resolved DOS for the AFM bilayer K$_2$CoS$_2$ in Figure \ref{fig4}(b) shows an insulating state with a large bandgap of $\sim$2.49 eV and VHSs similar to the bulk and monolayer K$_2$CoS$_2$.

\begin{table}[ht!]
\caption{Relaxed lattice parameters (in \AA) for bilayer K$_2$CoS$_2$ along with energies (E) per Co atom for various magnetic configurations}  
\centering  
\resizebox{0.99\columnwidth}{!}{
\begin{tabular}{ c c c c c }
\hline \hline
Configuration & \hspace{0.15cm} $a$ (\AA) \hspace{0.15cm} & \hspace{0.15cm}$b$ (\AA) \hspace{0.15cm}& \hspace{0.15cm}Area (\AA$^2$) \hspace{0.15cm} & \hspace{0.15cm} E (eV) \\ 
\hline \hline 
NM & 6.591 &  5.975  & 39.381 & -17.091\\
\hline
AFM & 6.743 & 6.113  & 41.219 & -19.160\\
\hline
FM & 6.732 & 6.103 & 41.085 & -18.563 \\
\hline
\end{tabular}}\label{tab3} 
\end{table}

\section{Conclusion} \label{conclusion}
We predict layered K$_2$CoS$_2$ as a new, thermodynamically stable 2D AFM material with a quasi-1D Neel ordering of the magnetic moments along the Co-chains. The interlayer exchange coupling is also shown to be antiferromagnetic, making K$_2$CoS$_2$ a unique system with robust in-plane AFM ordering from monolayer to bulk. Our systematic first-principles calculations reveal the presence of flat bands and VHSs around the Fermi energy in the nonmagnetic bulk as well as monolayer K$_2$CoS$_2$. Our classical Monte Carlo simulations on monolayer K$_2$CoS$_2$ predict its transition temperature to be $T_N \approx 15$ K, which is close to the experimentally observed value in bulk K$_2$CoS$_2$\cite{bronger1990}. Our study not only predicts a new 2D antiferromagnet with quasi-1D AFM chains but also provides a unique setting for exploring layer-dependent magnetism in K$_2$CoS$_2$ and related Co-based quasi-1D materials A$_2$CoB$_2$ (A $= \{$Na, Rb, and Cs$\}$ and B$=\{$S, Se$\}$).

\acknowledgements
We thank Sougata Mardanya for helpful discussions. A. B. S. acknowledges IIT Kanpur for providing Junior Research Fellowship. B. G. acknowledges CSIR-INDIA for the Senior Research Fellowship. We thank the CC-IITK for providing the HPC facility. A. A. acknowledges funding from Science education and research board (SERB) and Department of Science and Technology (DST), government of India. The work at Northeastern University was supported by the US Department of Energy (DOE), Office of Science, Basic Energy Sciences grant number DE-SC0019275 and benefited from Northeastern University's Advanced Scientific Computation Center (ASCC) and the NERSC supercomputing center through DOE grant number DE-AC02-05CH11231.

\appendix

\section{Estimating the value of effective onsite Hubbard U parameter}\label{hubbard}

\begin{figure}[h]
\includegraphics[width=0.9\linewidth]{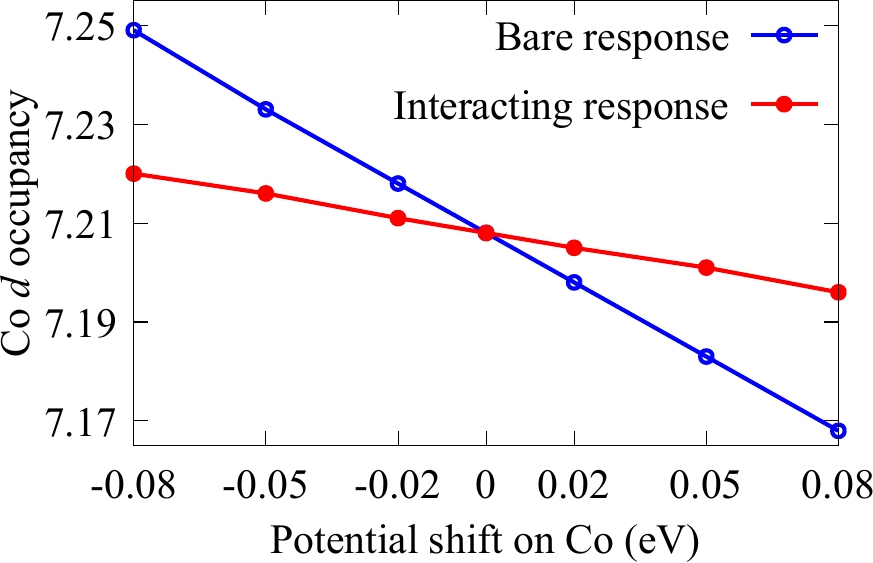} 
\caption{ Co-$d$ occupancy as a function of ‘potential shift’ on a Co atom in monolayer K$_2$CoS$_2$ as discussed in the text.}
\label{figS1}
\end{figure}

Value of the onsite Hubbard $U$ parameter for Co $d$ states was estimated by using the self-consistent method of Ref.~[\onlinecite{PhysRevB.71.035105}] based on the linear response theory. Following Ref.~[\onlinecite{PhysRevB.71.035105}], we calculated bare and interacting responses in terms of Co-$d$ occupancy (total number of occupied Co $d$ electrons) for various `potential shifts' applied on a Co site in monolayer K$_2$CoS$_2$. Variations in Co $d$ occupancy as a function of potential shift are seen from Fig.~\ref{figS1} to follow a linear relationship. Using the density response functions in Eq.~18 of Ref.~[\onlinecite{PhysRevB.71.035105}], yields a $U$ value of 4.4 eV for monolayer K$_2$CoS$_2$. Our estimate of $U$ is based on a single unit cell of 10 atoms. A larger supercell requiring more intense computations will be required for obtaining a more precise value of $U$. In any event, we have verified the robustness of our results by varying the value of $U$ from 0 to 6 eV. In particular, the AFM ground state is found to remain intact over a wide range of $U$ values.

\begin{table*}[t!]
	\caption{Relative energies per unit cell for various magnetic configurations of monolayer K$_2$CoS$_2$ defined as differences: E$^{x/y/z}_{AFM/FM}$ - E$^{y}_{AFM}$, where superscripts denote the direction of spin orientation. Results for various $U_{eff}$ (eV) values are given. }  
	\centering  
	\begin{tabular}{>{\centering\arraybackslash}p{2cm}>{\centering\arraybackslash}p{1.75cm}>{\centering\arraybackslash}p{1.75cm}>{\centering\arraybackslash}p{1.75cm}>{\centering\arraybackslash}p{1.75cm}>{\centering\arraybackslash}p{1.75cm}>{\centering\arraybackslash}p{1.75cm}>{\centering\arraybackslash}p{1.75cm}>{\centering\arraybackslash}p{1.75cm}}
		\hline \hline
		Configuration &  $U_{eff}$ = 0 & $U_{eff}$ = 1.0 & $U_{eff}$ = 2.0 & $U_{eff}$ = 3.0 & $U_{eff}$ = 4.0 & $U_{eff}$ = 4.4 & $U_{eff}$ = 5.0 & $U_{eff}$ = 6.0 \\ 
		\hline \hline
		AFM$_y$ &  0 & 0 & 0 & 0 & 0 & 0 & 0 & 0 \\
		\hline
		AFM$_x$ &  0.0004 & 0.0008 & 0.0009 & 0.0011 & 0.0011 & 0.0011 & 0.0009 & 0.0008 \\
		\hline
		AFM$_z$ &  0.0002 & 0.0007 & 0.0009 & 0.0011 & 0.0011 & 0.0011 & 0.0010 & 0.0009 \\
		\hline
		FM$_y$ &  -0.0074 & 0.1983 & 0.2174 & 0.1790 & 0.1408 & 0.1270 & 0.1082 & 0.4187 \\
		\hline
		FM$_x$ &  5.1720 & 0.2168 & 0.2250 & 0.1838 & 0.1443 & 0.1302 & 0.1110 & 0.2942 \\
		\hline
		FM$_z$ &  5.1719 & 0.2166 & 0.2250 & 0.1838 & 0.1444 & 0.1302 & 0.1111 & 0.2941 \\
		\hline \hline
	\end{tabular} 
	\label{tab1}
\end{table*}


\section{Energies of different magnetic configurations of monolayer $\mathbf {K_2CoS_2}$ for various values of $U$} \label{energy}

The calculated relative energies per unit cell for different magnetic configurations of monolayer K$_2$CoS$_2$ are listed in Table.~\ref{tab1} for various values of $U$. The Magnetic state is favoured over the nonmagnetic state in all cases. The AFM state with magnetic moments aligned along the Co chain ($y$-direction) remains the lowest energy state for all nonzero $U$ values. However, the negative value of relative energy for FM$_y$ state at $U=0$ suggests some tendency toward the FM$_y$ state.

\section{Values of parameters in the model Hamiltonian for monolayer K$_2$${\bf Co}$S$_2$ with 1st and 2nd nearest neighbour couplings}\label{anisotropy}

Our minimal model Hamiltonian with various magnetic couplings and onsite magnetic anisotropy is: 

\begin{widetext}
\begin{equation}\label{Ham}
	 H =  -\frac{1}{2} \sum_{\langle ij\rangle}\left[J^{x,1st}_{i,j}S^x_{i}S^x_{j} + J^{y,1st}_{i,j}S^y_{i}S^y_{j} + J^{z,1st}_{i,j}S^z_{i}S^z_{j} \right]  - \frac{1}{2} \sum_{\langle kl \rangle}\left[J^{x,2nd}_{k,l}S^x_{k}S^x_{l} + J^{y,2nd}_{k,l}S^y_{k}S^y_{l} + J^{z,2nd}_{k,l}S^z_{k}S^z_{l} \right] -  \sum_{i} D (S^y_i)^2~.
\end{equation}
\end{widetext}
Here, the summations $\langle ij \rangle$ and $\langle kl \rangle$ run over the $1^{st}$ and $2^{nd}$ nearest neighbour Co atoms, respectively. 

\begin{figure}[t]
\includegraphics[width=0.99\linewidth]{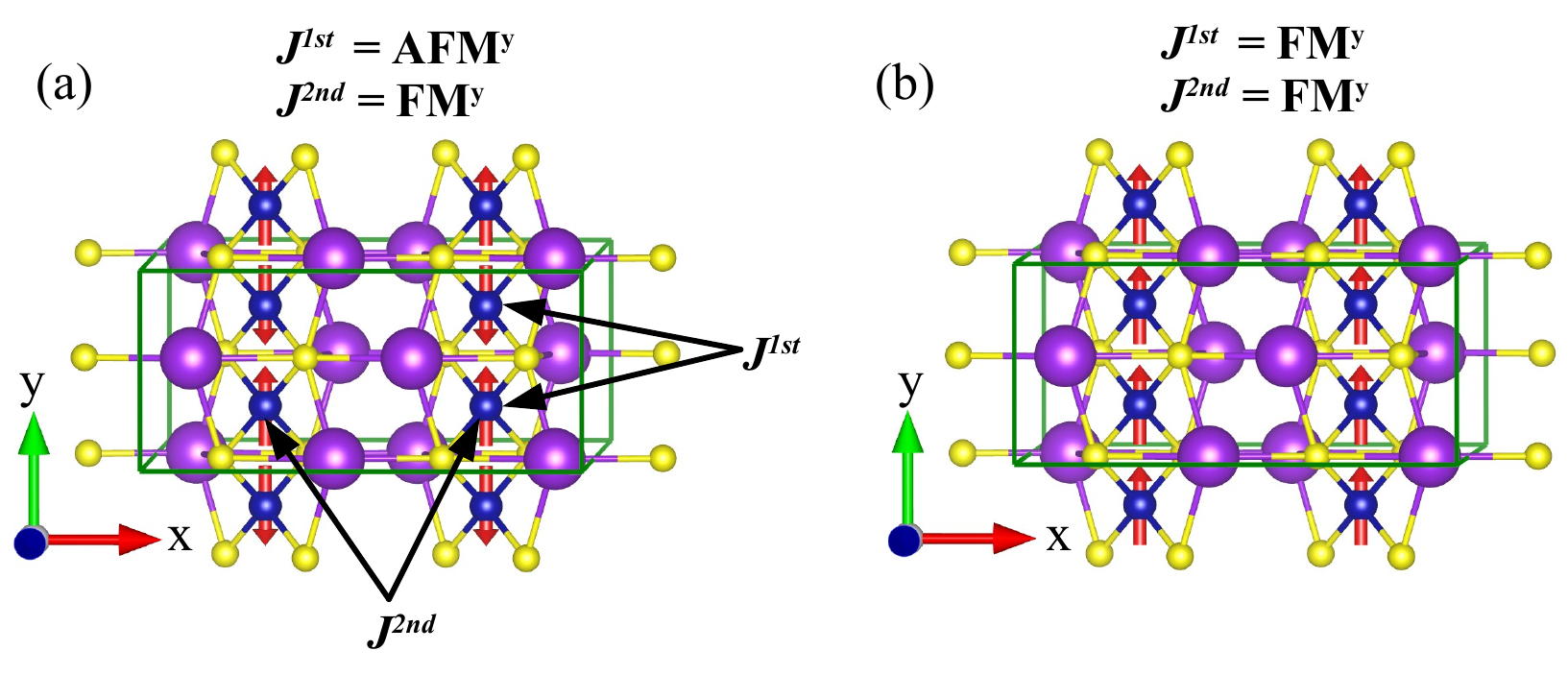} 
\caption{Monolayer K$_2$CoS$_2$ with intra-chain Co magnetic moments in (a) AFM  and (b) FM configuration. The $J^{1st}$ and  $J^{2nd}$ exchange interaction parameters along the intra- and inter-chain directions are marked. The unit cell is doubled along the $x$ axis to capture the interactions between the inter-chain Co atoms. Magnetic moments are oriented along the Co-chain ($y$) direction.}
\label{figs1}
\end{figure}

{\it Calculation of $J^{1st}$.} 
The intrachain exchange coupling parameter ($J^{1st}$) is obtained by employing the magnetic configuration given in 
Fig.~\ref{figs1} as: 
\begin{equation}
	J^{1st}_a = \frac{E^{AFM}_a - E^{FM}_a}{8S^2}~,
\end{equation}
where, $a$ = $\{x,y,z\}$ and $S$ is the magnitude of the magnetic moment. Note that the inter-chain Co moments are kept fixed in an FM configuration as shown in Fig. ~\ref{figs1}. 

\begin{figure}[t!]
	\includegraphics[width=0.99\linewidth]{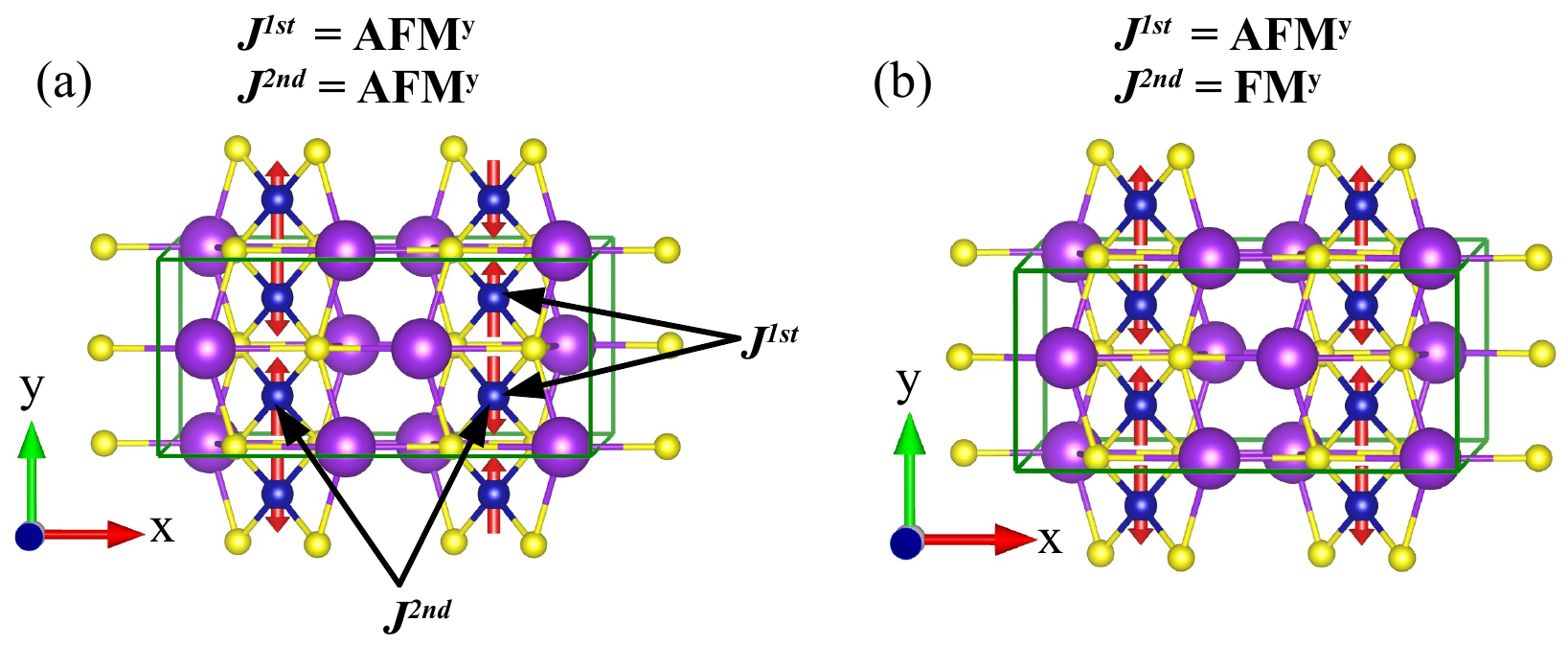} 
	\caption{Same as Fig.~\ref{figs1} except that here the interchain Co moments are set to (a) AFM and (b) FM configuration whereas the intrachain Co moments are fixed in AFM configuration.  }
	\label{figs2}
\end{figure}

{\it Calculation of $J^{2nd}$.}
The interchain exchange coupling parameter $J^{2nd}$ is obtained using the magnetic configuration given in Fig.~\ref{figs2} in a manner similar to that described above for $J^{1st}$. Notably, the intrachain magnetic moments are now fixed in the AFM configuration whereas the interchain magnetic moments are switched to AFM and FM configurations (see Fig.~\ref{figs2}). $J^{2nd}$ is defined as:

\begin{equation}
J^{2nd}_a = \frac{E^{AFM}_a - E^{FM}_a}{8S^2}~,
\end{equation} where $a$ = $\{x,y,z\}$ and $S$ is the magnitude of the magnetic moment. 

{\it Calculation of the anisotropy parameter $D$.}
The anisotropy parameter $D$ is calculated from the energy difference between the AFM configurations with magnetic moment aligned along the $x$ and $y$-directions. $D$ is defined as  
\begin{equation}
	D = \frac{(E^x_{AFM} - E^y_{AFM}) - (j^{1st}_x + j^{2nd}_x - j^{1st}_y - j^{2nd}_y)4S^2}{4S^2}~.
\end{equation}

\bibliography{Ref_Co_1}
\end{document}